# Analysis of HAT-P-23 b, Qatar-1 b, WASP-2 b, and WASP-33 b with an Optimized EXOplanet Transit Interpretation Code

*Sujay Nair*
*Stanford Online High School*
*Academy Hall Floor 2 8853, 415 Broadway Redwood City, CA 94063, USA*

*Jonathan Varghese*
*Vista Del Lago High School*
*1970 Broadstone Pkwy, Folsom, CA 95630, USA*

*Kalée Tock*
*Stanford Online High School*
*Academy Hall Floor 2 8853, 415 Broadway Redwood City, CA 94063, USA*

*Robert Zellem*
*Jet Propulsion Laboratory, California Institute of Technology 4800*
*Oak Grove Drive, Pasadena, California, 91109, USA*

**Abstract**

The ability for citizen scientists to analyze image data and search for exoplanets using images from small telescopes has the potential to greatly accelerate the search for exoplanets. Recent work on the Exoplanet Transit Interpretation Code (EXOTIC) enables the generation of high-quality light curves of exoplanet transits given such image data. However, on large image datasets, the photometric analysis of the data and fitting light curves can be a time-consuming process. In this work, we first optimize portions of the EXOTIC codebase to enable faster image processing and curve fitting. Specifically, we limited repetitive computation on fitting centroids with various apertures and annuli. Moreover, this speedup is scaled linearly based on the number of FITS files. After testing on existing HAT-P-32 b data and newer HAT-P-23 b data, our best demonstration was approximately a 5x speedup, though that factor increases given a larger number of FITS files. Utilizing the accelerated code, we analyzed transits of HAT-P-23 b, Qatar-1 b, WASP-2 b, and WASP-33 b using data captured by the 16" SRO telescope operated by Boyce-Astro.

## 1. Introduction

Exoplanets are planets outside of our solar system. Two common methods of discovering these planets are the transit method and the radial velocity method; however, in this work, we use the transit method. Specifically, by plotting the relative brightness of the star against a comparison star over an exoplanet's transit (when a planet passes in front of a star), in what is known as a light curve, a dip in flux can be measured, possibly indicating the existence of an exoplanet. A comparison star is used to ensure that atmospheric/external variability doesn't have an impact on the dip. The mid-transit time is the time in which the exoplanet is in the middle of its transit (tracked by the NASA Exoplanet Archive Akeson et al. (2013). The mid-transit time error increases over time due to the uncertainty in the period of the exoplanet. Because of this error, freshening transit midpoints, or continuously observing light curves of the exoplanet to see the transit midpoint, is necessary for accurate transit times (Zellem et al. 2020).

EXOTIC is a codebase that takes in FITS files or a pre-reduced text file and creates a light curve. After specifying the x and y pixels for the target and comp stars, EXOTIC runs its own photometric algorithm. From there, EXOTIC would run a Markov Chain Monte Carlo (MCMC) to best estimate the light curve parameters. The two most time-consuming portions were the MCMC and the photometry. To increase efficiency, edits were made to the photometric algorithm to avoid repetitive computation, and the MCMC was tested with multiple CPUs and a GPU on the Google Cloud Platform (GCP).

First, we will overview the targets selected, and data sources used. We chose to analyze four exoplanets HAT-P-23 b, Qatar-1 b, WASP-2 b, and WASP-33 b. Within our analysis of these planets, we worked with uncalibrated and calibrated data. Additionally, we worked with two planets containing a meridian flip, Qatar-1 b and WASP-2 b (a meridian flip occurs when the star crosses the meridian and the telescope mount needs to be rotated 180 degrees). We will then discuss the optimizations made to the EXOTIC photometry. Lastly, we will present the light curves and the estimated transit parameters of the analyzed planets.

## 2. Target Selection

We analyzed 4 exoplanets: HAT-P-23 b, Qatar1 b, WASP-2 b, and WASP-33 b. We chose to use these exoplanets because the data was previously unanalyzed. We used the uncalibrated data to test EXOTIC's use of calibration frames and to compare the difference in quality between the calibrated and uncalibrated light curves.

Within our observations, things we hoped for included optimal parameters for analysis using EXOTIC; specifically, we looked at high expected transit depths and short transit durations in addition to short periods; see Table 1 (at end of paper). We chose WASP-33 b to test if EXOTIC could plot an exoplanet without a given transit depth or midpoint in the NASA Exoplanet Archive Akeson et al. (2013).

We received our image data from Pat Boyce of the SRO Observatory. Sierra Remote Observatory (SRO) is located in the California Sierra Mountains at an elevation of around 1400m. Additionally we utilized Gaia data release 2 to see if comparison stars were variable and the NASA Exoplanet Archive Akeson et al. (2013) to compute phase differences and obtain various stellar/planet traits. See Table 2 for exposure settings.

Table 2. Exposure settings for exoplanets used

|  | Exposure Time (sec) | Binning |
|---|---|---|
| HAT-P-23 b | 30 | 3x3 |
| Qatar-1 b | 30 | 2x2 |
| WASP-2 b | 30 | 2x2 |
| WASP-33 b | 30 | 2x2 |

## 3. Methods

### 3.1 How Exotic Works

Information on running EXOTIC in more detail can be seen in Zellem et al. (2020). The GitHub can be seen here Rzellem (2020). To run the FITS files on EXOTIC, the data were first downloaded to a directory on the local machine only for consistency, though it is possible to run EXOTIC through Google Colab. If accessible, calibration files (darks, flats, and biases) would be applied to the FITS files to clear camera noise and create clearer images. Using a FITS file viewer, the image pixel coordinates (known as a centroid) of the target star, as well as the centroids of up to 10 comparison stars, were entered into EXOTIC. Out of the comparison stars inputted, EXOTIC selects the best comparison star based on the lowest residual scatter. EXOTIC then took the coordinates of the target and comparison star to fit an aperture and annulus for both stars. To ensure the centroid is always in the center of the star, EXOTIC would calculate the shift of the star throughout all the FITS files. Finally, EXOTIC plots the relative flux between the target star and comparison star to account for atmospheric fluctuations. The light curve is constructed based on these fluxes and uses four parameters: the mid-transit time, the ratio of the planet to the star, and 2 airmass constants. To find the correct parameters for the light curve, a Markov Chain Monte Carlo (MCMC) sampling method is used.

### 3.2 Running Exotic

For HAT-P-23 b, we were able to calibrate the data to make a calibrated light curve to compare our uncalibrated light curve to. We included the uncalibrated light curve to see the extent to which the calibration frames would improve the light curve and to see if EXOTIC would have similar predicted fluxes. We chose a specific set of darks, flats, and biases based on the exposure time and binning of our images. The uncalibrated light curve can be seen in Figure 1(a) and the calibrated light curve can be seen in Figure 1(b).

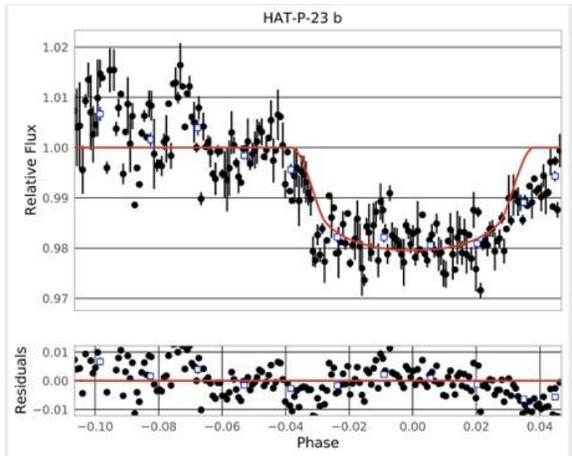

Figure 1s. Uncalibrated light curve of HAT-P-23 b

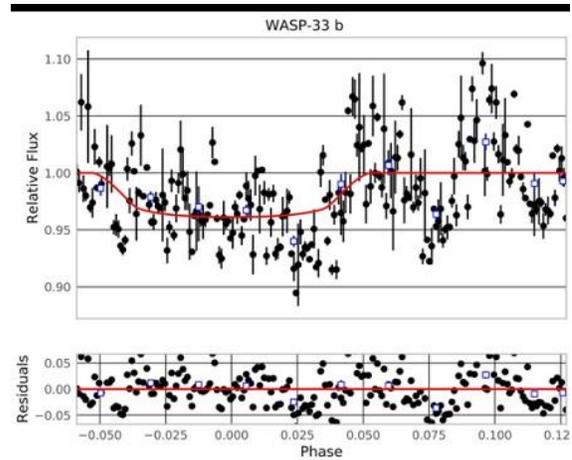

Figure 2b. The uncalibrated full transit. uncalibrated.

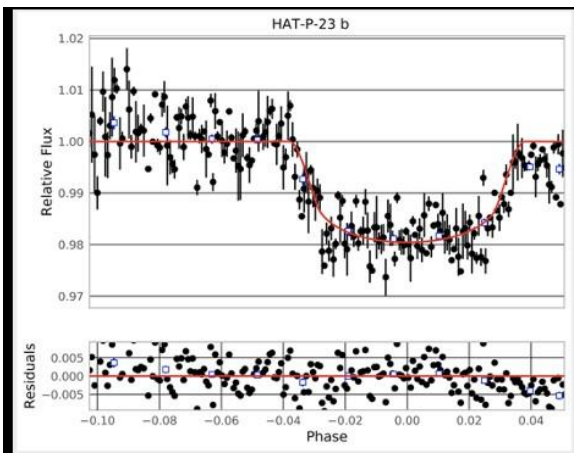

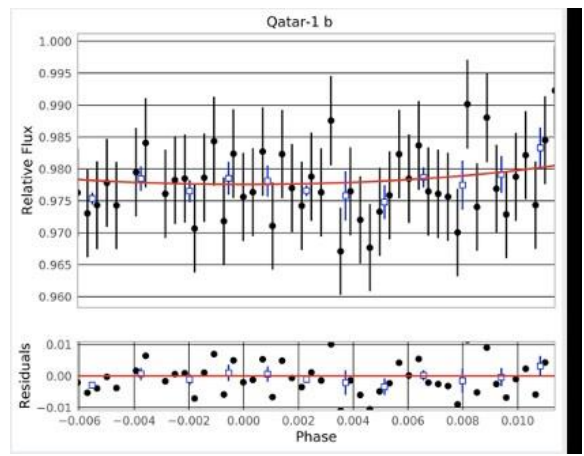

Figure 1b. Calibrated light curve of HAT-P-23 b. The calibrated data models the line of fit more clearly with Figure 3a. We see the uncalibrated light curve made significantly less scatter. from just a portion of the full transit.

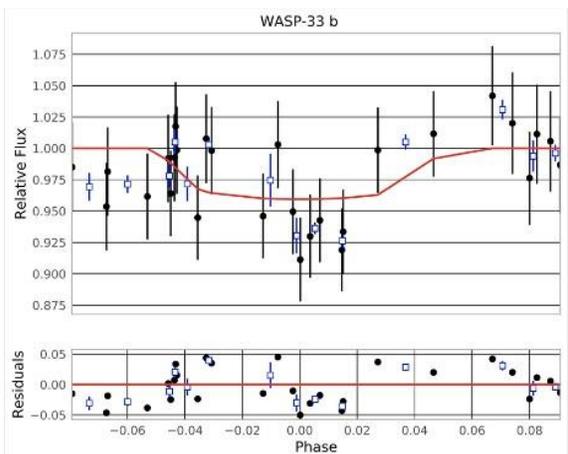

Figure 2a. The uncalibrated light curve of the smaller portion of the transit.

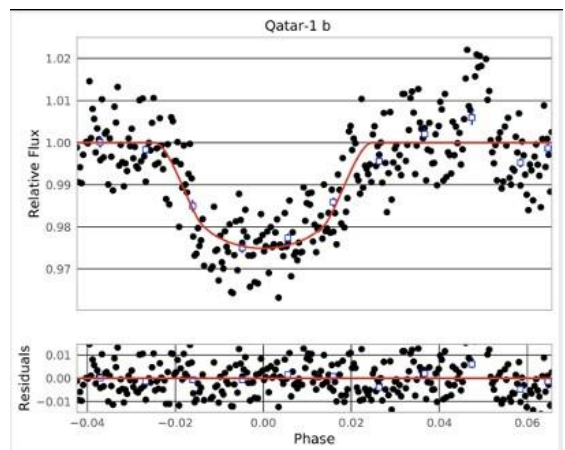

**Figure 3b.** Depicts the complete uncalibrated transit.

For WASP-33 b, when using the data initially, we only had a portion of it due to an error in downloading the files. We ran EXOTIC with what data we had in the beginning, as seen in Figure 2(a) and later downloaded and ran the full dataset seen in Figure 2(b).

For Qatar-1 b, we initially ran into the issue of only having a partial series of images like WASP-33 b; nevertheless, EXOTIC was still able to reduce that data and complete a light curve as seen in Figure 3(a). After sorting out data problems, we ran the full dataset, only to see that there was a meridian flip in the data. A meridian flip rotates a large part of the images causing EXOTIC to lose the location of the target/comparison star as the pixel coordinates of both will be very different after the flip. When this happens, EXOTIC runs on only the FITS files prior to the meridian flip, producing a partial transit light curve. To account for this, we ran all images before and after the flip separately such that EXOTIC would give us the text file with the normalized fluxes. Both files were then combined, and the combined file was then inputted into EXOTIC as pre-reduced data. This became what we see in Figure 3(b).

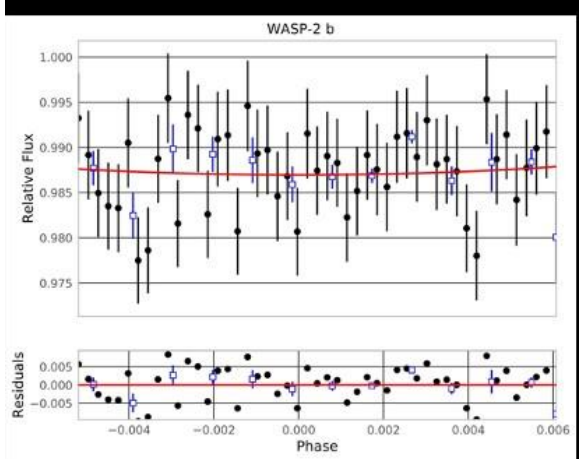

**Figure 4a. Here we see in the uncalibrated light curve of WASP-2 b till the meridian flip.**

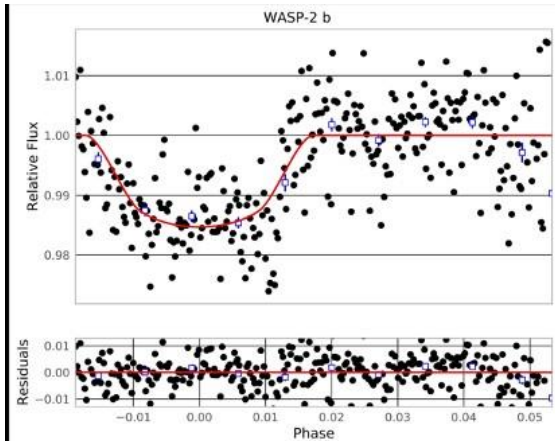

**Figure 4b. Here we see in the uncalibrated light curve beyond the meridian flip.**

WASP-2 b also had a meridian flip and similar to Qatar-1 b, we were able to depict the light curve from before the flip and from after separately as the first part is seen in Figure 4(a) and the second part is seen in Figure 4(b). Our comparison star parameters can be seen in Table 3 (at end of paper).

### 3.3 Speed Up of EXOTIC

Although given clean data EXOTIC will correctly interpret the transit, one drawback is how time-consuming it can be. Some of the most timeconsuming sections include the photometry aspect and the curve fitting with the Markov Chain Monte Carlo (MCMC). Part of this work includes speeding up the photometric reduction aspect and exploring ways of possibly making the MCMC faster using GPUs on the Google Cloud Platform (GCP).

#### 3.3.1. Centroid Fitting and Photometry

One part of the code that was time consuming was the centroid fitting loop. Once EXOTIC receives the initial pixel coordinates of the star on the FITS, it runs all combinations of apertures and annuli while adjusting the centroid coordinates to account for shifts of the star throughout the FITS files. The loop was nested such that EXOTIC would set one aperture and one annulus, then do calculations to adjust the centroid coordinates. The drawback was that it then would do these same calculations for every combination of aperture and annulus, which is unnecessary given that it was done on the first iteration for every image. We edited the codebase to store the coordinates of the centroid into a dictionary could be referred to for the following apertures and annuli, instead of having to recalculate the same adjustments every time.

The current build of EXOTIC has this photometry speedup implemented and specific speedups can be seen in Table 4 (at end of paper).

The updated speed is seen in Equation 1. This leads to a speedup of a factor seen in Equation 2. Also, this means the total speed improvement is Equation 3. The best speedup displayed in this work is approximately a factor of 5, though the speed up factor on the photometry increases linearly based on the number of apertures and annuli and the total speedup scales linearly with the number or images.

An advantage to this is the ability to now go through many more comparison stars quicker to have the best possible comparison star of many more options.

### 3.3.2. Speed Up Equations

For the following calculations, let N = the number of images, Ap = the number of apertures to be tested, An = the number of annuli to be tested, Tc = the time to fit the centroid, and Tr = the time to register a translation.

$$\text{Current Photometry Time} = N*Tr + 2Tc \quad (1)$$

$$\text{Photometry Factor Speedup} = Ap*An \quad (2)$$
$$\text{Photometry Speed Improvement}$$
$$= N*Tr + 2Tc*Ap*An - 1 \quad (3)$$

### 3.3.3. Markov Chain Monte-Carlo

One of the lengthiest and processor heavy components of EXOTIC is the MCMC fit. The MCMC randomly samples to obtain the best value for the mid-transit time, the ratio of planet to star, and 2 airmass constants. These four parameters are the basis for constructing a light curve. To obtain these parameters, the MCMC algorithm randomly chooses a mid-transit time, ratio of planet to star, and 2 airmasses, and then calculates the least squares residual value with this combination of parameters. It then selects a large, predetermined number of random combinations to eventually find the combination that gives the smallest error. It randomly jumps between parameter sets with probability $p$, where $p$ is how well the params fit the data.

To speed up this process, we attempted to use multiple CPUs and a P100 GPU, all of which were on a GCP Virtual Machine instance. Overall the CPUs didn't have a significant impact, likely due to the fact that the sampling caps out at a max of 4 cores. We thought using a GPU would speed up the process because it seemed that the MCMC could be easily parallelized, especially considering that it used Theano tensors which are GPU compatible. However, the GPU did not have a significant improvement and slowed down the MCMC in some instances as seen in Table 5. When testing the GPU versus the CPU on the cloud, the number of cores used when sampling was fixed to 1 to avoid some issues the GPU would have when increasing the number of cores. Though, Table 5 clearly shows that the GPU was not significantly better, in fact it appears it was worse, than the CPU, indicating that one should generally stick to regular CPUs when running the MCMC in EXOTIC.

## 4. Observations

One of the lengthiest and processor heavy components of EXOTIC is the MCMC fit. The MCMC randomly samples to obtain the best value for the mid-transit time, the ratio of planet to star, and 2 airmass constants. These four parameters are the basis for constructing a light curve. To obtain these parameters, the MCMC algorithm randomly chooses a mid-transit time, ratio of planet to star, and 2 airmasses, and then calculates the least squares residual value with this combination of parameters. It then selects a large, predetermined number of random combinations to eventually find the combination that gives the smallest error. It randomly jumps between parameter sets with probability $p$, where $p$ is how well the params fit the data.

To speed up this process, we attempted to use multiple CPUs and a P100 GPU, all of which were on a GCP Virtual Machine instance. Overall the CPUs didn't have a significant impact, likely due to the fact that the sampling caps out at a max of 4 cores. We thought using a GPU would speed up the process because it seemed that the MCMC could be easily parallelized, especially considering that it used Theano tensors which are GPU compatible. However, the GPU did not have a significant improvement and slowed down the MCMC in some instances as seen in Table 5. When testing the GPU versus the CPU on the cloud, the number of cores used when sampling was fixed to 1 to avoid some issues the GPU would have when increasing the number of cores. Though, Table 5 (at end of paper) clearly shows that the GPU was not significantly better, in fact it appears it was worse, than the CPU, indicating that one should generally stick to regular CPUs when running the MCMC in EXOTIC.

### 4.1 Observations of HAT-P-23 b

During our analysis of HAT-P-23 b, light curves were made of raw/uncalibrated data, and calibrated data using darks/flats/bias. Figure 1(a) shows the light curve without any additional calibration frames whereas Figure 1(b) shows the light curve including calibration frames. Visually, we can see EXOTIC was able to fit a curve with less scatter in general with the calibrations. This is confirmed as we see in Table 6 that both the transit depth uncertainty and the scatter in the residuals are significantly less than that of the uncalibrated light curve.

Additionally, it appears that both light curves have stale transit midpoints due to the midpoint being slightly off to the right rather than in the center. However, when compared to an existing confirmed transit midpoint of 2454852.26599 (BJD), we see a phase difference of 23.871 minutes indicating that the midpoint was slightly stale. See Table 7 (at end of paper).

| HAT-P-23 b | Transit Depth Uncertainty (%) | Scatter in Residuals (%) |
|---|---|---|
| Uncalibrated | 0.1558 | 0.612 |
| Calibrated | 0.1427 | 0.464 |

Table 6. Difference between calibrated and uncalibrated data for HAT-P-23 b. The calibrated data produced a higher quality light curve as seen in the lower residuals and uncertainty values.

### 4.2 Observations of WASP-33 b

Throughout our analysis of WASP-33 b we made light curves of a partial series of raw images, and the full set of raw images. Our partial transit light curve is depicted in Figure 2(a) whereas our full transit light curve is depicted by Figure 2(b). Additionally, it was interesting to see that even with a partial series of images, EXOTIC was able to have a decent estimate for the transit midpoint and depth, while also capturing the transit duration well.

Phase difference in mid-transit = ((EXOTIC mid-transit – Expected mid-transit)/Period) (Mod 1) 
(4)

EXOTIC's mid-transit time was 2458027.7444759 whereas a previous mid-transit time was 2454590.17948. This yields a phase difference of over 1000 minutes which is very odd. It's likely that this large difference was either due to noise in the data or a misfit by EXOTIC. See Table 7.

### 4.3 Observations of Qatar-1 b

Over the course of our analysis of the Qatar-1 b, light curves were made with the raw, uncalibrated fits files, calibrated data, data containing a meridian flip, and partial transits. Figure 3(a) shows a 50-image section of the transit near the transit midpoint. This plot was made due to initial missing data, though the transit depth seems quite close to the expected 2.14% as seen in Table 1. This leads us to believe that EXOTIC correctly fit the data although there was a substantial amount of noise.

The final complete raw data light curve can be seen in Figure 3(b). It's likely that the fit is accurate as the transit depth from EXOTIC was 2.3% which was very similar to the expected 2.14% in addition to this data being uncalibrated. On the note of using darks/flats/bias to calibrate these images, the provided calibration frames were offset and threw an error with EXOTIC at the point of the meridian flip, and so the fluxes in the light curve after the flip were incorrect. Since we couldn't obtain any normalized flux values from the calibrations on both sides of the flip (as the calibrations were not impacted by the flip), we couldn't use them.

EXOTIC's mid-transit time was 2457960.85276157 whereas a previous mid-transit time was 2456234.10322. This yields a phase difference of close to 40 minutes. Like HAT-P-3 b, this was slightly stale. See Table 7.

### 4.4 Observations of WASP-2 b

During our analysis of WASP-2 b, we looked at data before and after the meridian flip. Since we couldn't use text files containing the normalized flux from before and after the flip due to errors with comparison stars not being measured properly, we just plotted both sides separately. Figure 4(a) illustrates the light curve until the 40th image whereas Figure 4(b) illustrates the light curve from the 40th image to the end.

EXOTIC's mid-transit time was 2457955.9070491 whereas a previous mid-transit time was 2458339.00342. This yields a phase difference of over 2500 minutes. Again, it's very likely that this large difference was either due to noise in the data or a misfit by EXOTIC. See Table 7.

### 4.5 Summary of Results

On the note of freshening mid-transit times, we saw that the mid-transit phase differences for HAT-P23 b and Qatar-1 b were 20 minutes and 40 minutes respectively. WASP-2 b and WASP-33 b both had extremely high differences pointing toward a miscalculation by EXOTIC or noise in the data. These values can be seen in the first 4 columns of Table 7.

Overall, EXOTIC's transit depth predictions were pretty accurate as the average depth difference between HAT-P-23 b, Qatar-1 b, and WASP-2 b was approximately 0.193%. There was, however, a substantial difference between the EXOTIC transit depth and the expected transit depth for WASP-33 b as seen in the final column of Table 7. Although the transit depth was overestimated by EXOTIC, the light curve in Figure 2(b) does display a clear dip. This dip may have been incorrect however, possibly linked to the strange mid-transit phase difference.

## 5. Conclusion

We were able to successfully produce a calibrated light curve for HAT-P-23 b and saw a clear decrease in uncertainty and residuals using calibration frames, although the meridian flip in our Qatar-1 b and WASP-2 b data created errors when running the calibrations. It appears that using calibration frames with EXOTIC significantly reduces transit depth uncertainty and scatter in the residuals. To run Qatar-1 b data uncalibrated, the data before and after the flip were separated, converted to pre-reduced text files and fed back into EXOTIC, which worked well enough to produce fairly good light curves. By combining the pre-flip and post-flip text files, we were able to produce a full light curve for Qatar-1 b. In the case of WASP-2 b, the combined text file was not producing a proper curve, so we had to only use the data before and after the meridian flip to create two light curves of partial transits. Finally, although running the MCMC using a GPU didn't significantly speed up the code, rewriting portions of the EXOTIC program to avoid unnecessary recalculations of the centroid adjustments decreased the total run time of the photometry linearly.

Based on our measurements, HAT-P-23 b and Qatar-1 b had their mid-transit times freshened by 20 minutes and 40 minutes respectively. WASP-33 b and WASP-2 b had extremely large phase differences, and their light curves were quite noisy and were strange during the fitting process. Based on that, it's possible that the calculated phase differences for them are incorrect so we will say that their freshened transit properties are unclear.

## 6. Future Work

Due to the meridian flip of the FITS files of Qatar-1 b and WASP-2 b, EXOTIC could not create accurate light curves without having to split the data apart into before and after the meridian flip. This created complications in producing the light curves especially for the calibrated light curves. We would like to add code to the centroid fitting section in EXOTIC that could account for data with meridian flips, and alter the pixel coordinates depending on the degree of the rotation of the images. We could possibly look into plate solving the images first also.

The MCMC sampling still takes up a large portion of the EXOTIC run time due to the huge number of samples and the completely random selection. We would like to experiment with a gradient descent algorithm that, instead of randomly selecting parameters, would create a function that outputs the error of a given combination, and slowly adjust the parameters to arrive at a global minimum of this error function. This way, instead of having to always select a set number of combinations and run the residual calculations every time, the gradient descent algorithm would be able to find the smallest residual much faster by learning from the residual values of previous parameter combinations

We had trouble producing quality light curves with the flats, darks, and bias calibrations for all of our exoplanets except for HAT-P-23 b. These calibrations seemed to have increased the quality of the light curve when we were successful in running the data calibrated, however we were not able to run Qatar-1 b and WASP-2 b due to persistent errors. We hope that we will be able to solve these errors and run these calibrations successfully in the future.

We also would like to obtain more pre-reduced text files from an external pipeline to observe any increase or decrease in the quality of the light curves. These files differ from the FITS files because they contain information for a specific photometry method. EXOTIC has its own photometry method that is run and is what was used to produce all the light curves shown. It would be interesting to see if the various photometry have any effect on the quality of the light curves for these data.

## 7. Acknowledgements

This paper has made use of the NASA Exoplanet Archive, which is operated by the California Institute of Technology, under contract with the National Aeronautics and Space Administration under the Exoplanet Exploration Program.

This work has made use of data from the European Space Agency (ESA) mission GAIA (https://www.cosmos.esa.int/gaia), processed by the GAIA Data Processing and Analysis Consortium (DPAC, https://www.cosmos.esa.int/web/gaia/dpac/consortium). Funding for the DPAC has been provided by national institutions in particular the institutions participating in the GAIA Multilateral Agreement.

This work has made use of data from the Sierra Remote Observatory (SRO) (https://www.sierraremote.com/).

This publication makes use of data products from Exoplanet Watch, a citizen science project managed by NASA's Jet Propulsion Laboratory on behalf of NASA's Universe of Learning. This work is supported by NASA under award number NNX16AC65A to the Space Telescope Science Institute.

## 8. References

Akeson, R. L., Chen, X., Ciardi, D., et al. (2013), *Pub. Astron. Soc. Pac.* **125**, 989

Bakos, G., Hartman, J., Torres, G., et al. (2011), *Astrophysical J.*, **742**, 116, doi: 10.1088/0004637x/742/2/116
*Astron. & Astrophys*, **553**, A44, doi: 10.1051/0004-6361/201219642

Ciceri, S., Mancini, L., Southworth, J., et al. (2015), *Astron. & Astrophys.*, 577, A54, doi: 10.1051/00046361/201425449

Kov´acs, G., Kova´cs, T., Hartman, J. D., et al. (2013),

Rzellem. (2020), rzellem/EXOTIC. https://github.com/rzellem/EXOTIC

Smith, A. M. S., et al. (2011), *Mon. Not. Royal Astro. Soc.*, **416**, 2096, doi: 10.1111/j.1365-2966.2011.19187.x

Zellem, R. T., Pearson, K. A., Blaser, E., et al. "Utilizing Small Telescopes Operated by Citizen Scientists for Transiting Exoplanet Follow-up." (2020) https://arxiv.org/abs/2003.09046

|  | Orbital Period (days) | Transit Depth (%) | Transit Duration (days) | Transit Midpoint (BJD) | Stellar Mag | Stellar Temp (K) |
|---|---|---|---|---|---|---|
| HAT-P-23 b | 1.21 | 1.4 | 0.091 | 2454852.26599 | 12.432 | 5905 |
| Qatar-1 b | 1.42 | 2.14 | 0.069 | 2456234.10322 | 12.843 | 5013 |
| WASP-2 b | 2.15 | 1.65 | 0.075 | 2458339.00342 | 11.88 | 5180 |
| WASP-33 b | 1.22 | 1.04 | 0.119 | 2454590.17948 | 8.142 | 7430 |

**Table 1.** Planet and Stellar parameters were stripped from the NASA Exoplanet Archive Akeson et al. (2013a). Values in yellow were not listed in the NASA Exoplanet Archive, though were found using other resources; specifically, the HAT-P-23 b mid-transit was found in Physical properties of the HAT-P-23 and WASP-48 planetary systems from multicolor photometry Ciceri, S. et al. (2015), the HAT-P-23 b transit duration was found in HAT-P-20b–HAT-P-23 b: FOUR MASSIVE TRANSITING EXTRASOLAR PLANETS Bakos et al. (2011), the WASP-33 b mid transit time was found using Thermal emission from WASP-33 b, the hottest known planet Smith et al. (2011), and finally the WASP-33 b transit depth was found in Comprehensive time series analysis of the transiting extrasolar planet WASP-33 b Kova´cs et al. (2013).

| Target | Comp RA HH:MM:SS | Comp Dec DD:MM:SS | Comp Parallax (mas) | Comp Magnitude | Comp Temperature (K) |
|---|---|---|---|---|---|
| HAT-P-23b | 20:24:41.54 | +16:46:28:15 | 1.7740 | 11.6800 | 6838.00 |
| Qatar-8b | 20:13:41.27 | +65:11:32.69 | 0.7273 | 12.7275 | 5085.71 |
| WASP-2b | 20:31:22.16 | +06:26:27.55 | 6.9702 | 11.7037 | 4896.73 |
| WASP-33b | 02:27:34.79 | +37:28:08.30 | 1.5322 | 9.2466 | 4993.64 |

**Table 3.** Information on best comparison stars for each target. Note that the best comparison star was the comparison that yielded the least residual scatter.

| Exoplanet | Number of .FITS files | Original Photometry Time (Seconds) | Edited Photometry Time (Seconds) | Comparison Stars Used |
|---|---|---|---|---|
| HAT-P-32b (sample data) | 142 | 29.92597 | 8.76661 | 2 |
| HAT-P-23b | 420 | 230.9185 | 52.1714 | 2 |

Table 4. Comparisons between the times of the original photometry and our edited version.

| Sample HAT-P-32b Data | Average Iterations/Second, MCMC Progress: 25% | Average Iterations/Second, MCMC Progress: 50% | Average Iterations/Second, MCMC Progress: 99% |
|---|---|---|---|
| GPU | 31.50 | 31.88 | 32.10 |
| CPU | 33.84 | 32.63 | 33.68 |

Table 5. Average Iterations per second at different points of the MCMC for both the GPU and CPU.

| | EXOTIC Mid-Transit Time (BJD) | Expected Mid-Transit Time (BJD) | Mid-Transit Observed vs Expected Phase Difference | Mid-Transit Observed vs Expected Phase Difference (Minutes) | EXOTIC Transit Depth (%) | Expected Transit Depth (%) | Transit Depth Difference (%) |
|---|---|---|---|---|---|---|---|
| HAT-P-23b | 2457920.8690187 | 2454852.26599 | 0.0137 | 23.871 | 1.62 | 1.40 | 0.22 |
| Qatar-1b | 2457960.8527616 | 2456234.10322 | 0.0208 | 42.551 | 2.38 | 2.14 | 0.24 |
| WASP-2b | 2457955.9070491 | 2458339.00342 | 0.8156 | ~2541 | 1.53 | 1.65 | 0.12 |
| WASP-33b | 2458027.7444759 | 2454590.17948 | 0.6762 | ~1187 | 3.33 | 1.04 | 2.29 |

Table 7. Differences in major parameters between EXOTIC and what was expected. The phase differences in red may have been due to noise or an incorrect fit as obtaining that numerical value is unlikely.